# Carbon cluster diagnostics-I: Direct Recoil Spectroscopy (DRS) of Ar⁺ and Kr⁺ bombarded graphite

Shoaib Ahmad[1,2,*], M.N. Akhtar[1], A. Qayyum[1], Bashir Ahmad[1], Khalid Bahar[1], Waheed Arshed[1]

[1]Accelerator Laboratory, PINSTECH, P.O. Nilore, Islamabad, Pakistan
[2]National Centre for Physics, Quaid-i-Azam University Campus, Shahdara Valley, Islamabad, 44000, Pakistan

Email: sahmad.ncp@gmail.com

**Abstract**

Measurements of the energy spectra of multiply charged positive and negative carbon ions ($C^{n\pm}$) recoiling from graphite surface under 100 and 150 keV argon and krypton ion bombardment are presented. With the energy spectrometer set at recoil angle of 79.5°, direct recoil (DR) peaks have been observed with singly as well as multiply charged carbon ions $C^{n\pm}$, where $n = 1$ to 6. These $C^{n\pm}$ ions have been observed recoiling with the characteristic recoil energy $E_{DR} = kE_0 \cos^2 \theta_{DR}$ where $\theta_{DR}$ is the direct recoil angle, $k = 4m_1 m_2 /(m_1 + m_2)^2$, $m_1$ and $E_0$ are projectile mass and energy and $m_2$ is carbon mass. We have observed sharp DR peaks. A collimated projectile beam with divergence $\sim \pm 0.2°$ is supplemented with a similar collimation before the energy analyzer to reduce the background of sputtered ions due to scattered projectiles.

## 1. Introduction

In our present investigations high angle charged recoils from graphite surface under heavy ion bombardments form the basis of Direct Recoil (DR) detection. In these experiments 100 and 150 keV Ar⁺ and Kr⁺ ions are incident on amorphous graphite. The technique used for charged particle detection is Direct Recoil Spectroscopy (DRS) which is the same as Elastic Recoil Detection (ERD) as developed by a number of investigators [1-4]. In ERD, generally smaller recoil angles $\sim$ 20-30° have been preferred for the detection of lighter impurities ($Z \leq 15$). At these angles heavy projectiles with 100 s of MeV (i.e., with energy $\sim$ 1 MeV /amu) result in lighter particles recoiling with at least a few MeV. Whereas, in case of DR studies [5-8,10] keV projectiles are used and recoil angles are generally large when scattered projectiles are to be avoided.

Regarding the charge state of direct recoils, ERD is insensitive to individual recoil's charge identification since it is mainly a particle detection technique and the detector collects neutral as well as charged species. DRS on the other hand, does provide a tool to distinguish between neutral and charged recoils and indicate the charge state as well [7,8]. Datz and Snoek [9] showed conclusively that solid surfaces could be used for studying individual collisions under energetic ionic bombardment and the ensuing mechanisms of charge transfer between projectiles and the target atoms.

Among the lighter targets, carbon in the form of amorphous graphite is a reasonable choice for isotropic materials. Graphite has been chosen for our present work because of its extensive use in reactors, fission as well as fusion.

The target atoms recoiling after elastic collisions with projectiles carry with them energies which are function of target to projectile mass ratio ($m_2/m_1$), angle of recoil $\theta_r$, and the bombarding energy $E_0$ The probability of occur-rence of a recoil in a particular direction has to be worked out from the calculations and measurements of differential cross sections $d u$ of these events. The cross sections for a recoil to occur for a particular combination of ($m_2/m_1$), $\theta_r$, and the bombarding energy $E_0$ can be worked out from the LSS [11,12] theory. However, the probability of a specific charge state of a recoil depends on factors like the type of chemical bonding of targets, formation of molecular orbitals during collisions [13,14] and the state of sputtering surface e.g., adsorption, ion induced roughing, etc. There have been attempts to provide explanations on the basis of models of Hagstrum [15] and Van der Weg and Pol [16].

Our choice of carbon as target and $Ar^+$ and $Kr^+$ as bombarding ions has been prompted by the requirements of moderate recoil energies in the range $0 < E_r < 5$ keV for an effective utilization of a compact 90° electrostatic energy analyzer (EEA) and secondly, the requirement to keep the data acquisition time per run ~ 20-30 min with a few μA beam. These conditions require that collision cross sections be such that recoil emissions at $\theta_r = 79.5°$ be significant and measurable.

The $E_{DR}$ measurements are performed at a fixed recoil angle $\theta_{DR} = 79.5°$. The energy of graphite atoms/clusters of mass $m_2$ recoiling at angle $\theta_{DR}$ is given by

$$E_{DR} = 4\{(m_1 m_2)/(m_1 + m_2)^2\} E_0 \cos^2 \theta_{DR} \qquad (1)$$

where $m_1$ and $E_0$ are mass and energy of projectile. For a given $E_0$ the cross section $d\sigma(E_0, E_r)$ for an energy transfer $E_r$, to the target atoms varies inversely as $E_r$ according to Sigmund [12] utilizing the LSS theory as

$$d\sigma(E_0, E_r) \propto E_0^{-m} E_r^{-1-m} dE_r \qquad (2)$$

where $m$ is a constant appropriately chosen for a given interaction. This cross section for $Ar^+$ incident on C with $E_0$ up to 200 keV for various recoil angles show that increasing $\theta$, from 40° to 60° and to 80° at a given $E_0$, $d\sigma(E_0, E_r)$ increases by more than an order of magnitude for 80° compared with that for 60°. This ratio is further enhanced when we go from $Ar^+$ to $Kr^+$. In our setup due to high recoil angle $\theta_r$, the scattered projectiles are avoided and the characteristic DR peaks are significantly measurable.

Eckstein [5] has conducted extensive research into low energy direct recoil studies of various targets including their work on graphite [6]. Their work is at lower values of $E_0$ which is an order of magnitude lower than our range of energies.

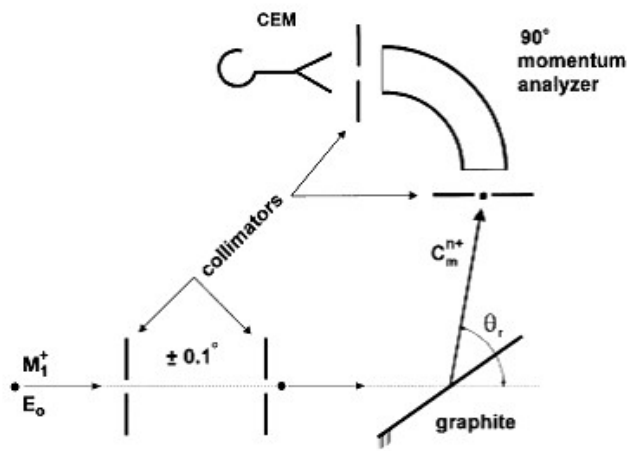

**Fig. 1.** The experimental setup for measurements of recoil spectra with projectile and recoiling species, the collimators, energy analyzer and the channel electron multiplier (CEM).

## 2. The experiment

### 2.1 The equipment

An indigenously designed and fabricated **PINSTECH** ion accelerator, a 250 keV heavy ion facility has been used for the experiments. $Ar^+$ and $Kr^+$ beams of > I mm diameter and energy between 50 to 250 keV can be delivered to a target 2 m from the end of the accelerator tube. The facility is equipped with a hollow cathode duoplasmatron [17] operating at $10^{-2}$ -$10^{-3}$ mbars with the accelerator delivering a few µA collimated beam with ∼ ±0.1° divergence on the target. The experimental setup is shown in Fig. 1, where the beam as well as the recoil particles' collimators with ± 0.1° divergence are shown along with a 90° electrostatic energy analyzer (EEA) and a channel electron multiplier (CEM). The resolution of the EEA is ∼0.02 with 0.8 mm entrance aperture for EEA. CEM with a typical gain of 5 x $10^7$ feeds the charged recoil data to a PC via a rate meter and a Hydra Data Acquisition unit. The EEA condenser plate potential is increased in variable steps through a function generator (Philips **PM** 5138). Solid angle of 8.7x$10^{-6}$ st. rad. is a compromise between high enough resolution and a decent count rate.

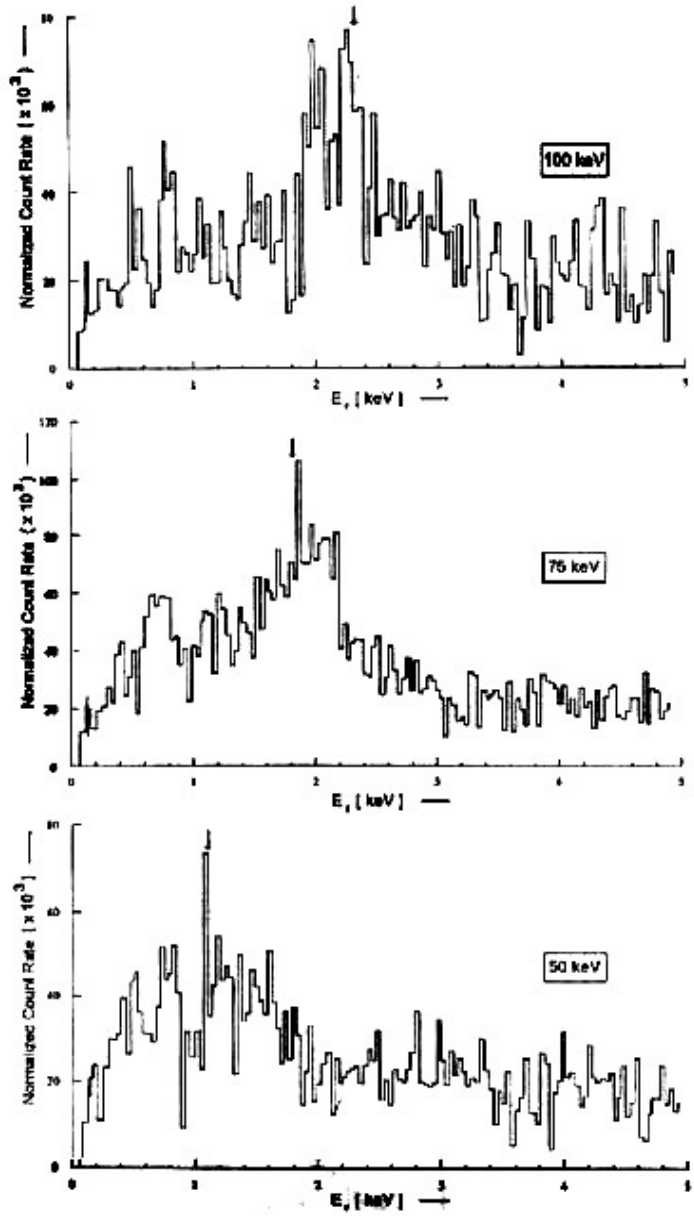

**Fig. 2** for $Ar^+ \rightarrow C$ the normalized count rate is plotted against the recoil energy $E_r$ for three projectile energies $E_0$= 100, 75 and 50 keV. Beam to target angle $\alpha$=15. Arrows point to the positions of $C_1$ direct recoils.

## 2.2. Experimental procedure

Experiments are performed at pressures $\sim 10^{-6}$ mbar at which the background gases have a tendency to adsorb at the surface. In addition, the starting surface is covered with many layers of $N_2$, $O_2$ and $H_2O$ molecules as our initial recoil spectra indicate. Few minutes of heavy ion bombardment at large angle to the surface normal i.e., $\alpha \sim 75°$ removes these adsorbed layers and gradually, the carbon surface starts to contribute predominantly in the energy spectra. During this initial cleaning process, secondary ions emitted are continually monitored and a systematic evaluation of adsorbed gases' contribution is done. A state of dynamic cleaning is achieved when $C^{n\pm}$ peaks dominate over those due to $N\pm$, $O\pm$ etc. For example, the dynamically clean surface at 100 keV $Kr^+$-C the ratio $C^+/N^+ \sim 4$ when counts per unit interval of recoil energy (counts/$\delta E$) are compared. The measured ratio of total counts $C^+/N^+ \sim 8$ and similarly for $C^-/N^- \sim 10$. Al-though the spectrum has large number of peaks but the cumulative $C^{n\pm}$ counts predominate over all other species. These numbers also indicate that less than a monolayer is expected to be on the surface during the experiments.

There is a peculiar advantage of having the adsorbed species present in the recoil spectra as we can calibrate the $E_{DR}^0$ pointers for $N^\pm$, $O^\pm$, etc., which help us to clearly distinguish various peaks due to $C^{n\pm}$. $N^\pm$ peaks are generally very sharp and help to establish positions of other peaks and act as markers in respective spectra.

## 3. Experimental results

Fig. 2 shows the recoil spectra of positively charged carbon ions sputtered at 79.5° by $Ar^+$ ions at 50, 75 and 100 keV, respectively. For all of the three bombarding energies, arrows point to the positions of expected DR peaks for singly charged recoils. The spread around the main peaks is noticeable. Also identifiable are the higher charge states at $E_{DR}/n$, where $n \geq 2$. It may be pointed out that the maximum recoil energy delivered to any mass in a binary collision at $\theta_{DR} = 79.5°$ is 3.32 keV i.e., for $(m_2/m_1) = 1$ and $E_0 = 100$ keV. Since we can see recoils up to S keV even in SO keV spectrum, which indicates a large contribution by the scattered projectiles. This contribution is seen on the higher side of the DR peaks with $E_r > E_{DR}$ which implies $\theta_r < \theta_{DR}$.

This particular set of data (Fig. 2) was taken with a single aperture of 0.8 mm diameter in front of the EEA. Although this subtended a small solid angle of 8.7 x $10^{-6}$ st. rad. but did allow $\delta\theta_r \sim \pm 0.6°$ which can introduce large errors in $\theta_{DR}$ and $E_{DR}$ measurements. This spread in recoil angle is due to the beam spread at large $\alpha$ angle. Rest of the data presented in Figs. 3-6 has been taken with a collimator placed in front of the EEA which limits the spread in $\delta\theta_r \sim \pm 0.1°$. The most important effect of the collimator in front of EEA is the reduction of scattered projectiles' contribution in the energy spectra which in turn reduces the background and enhances direct recoil peaks.

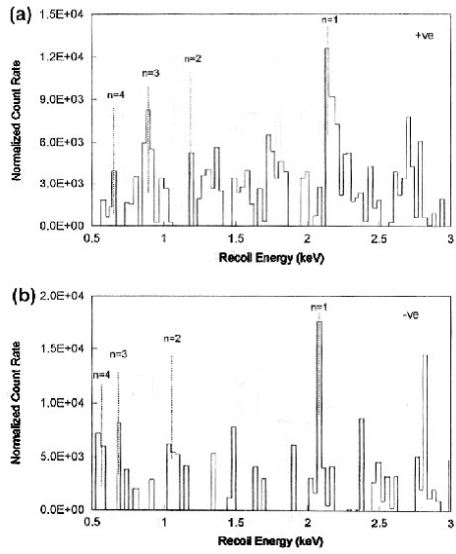

**Fig. 3.** The carbon recoil spectra is presented for $Kr^+ \rightarrow C$ at $E_0 = 150$ keV. (a) Positive recoils can be seen with respective charge states $n = 1$ to 6. (b) The negative species seem to have relatively clearer peaks.

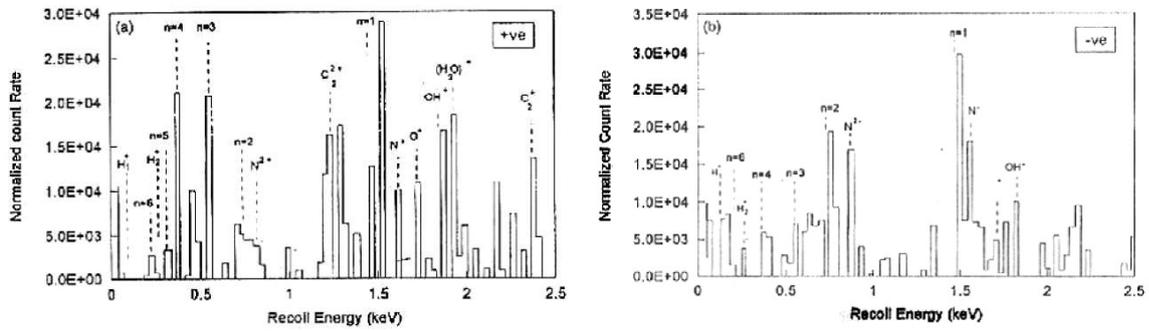

**Fig. 4.** $Kr^+ \rightarrow C$ at $E_0 = 100$ keV (a) $C^{n+}$ spectra along with the peaks for other species are shown. (b) The same but for the negative recoils $C^{n-}$.

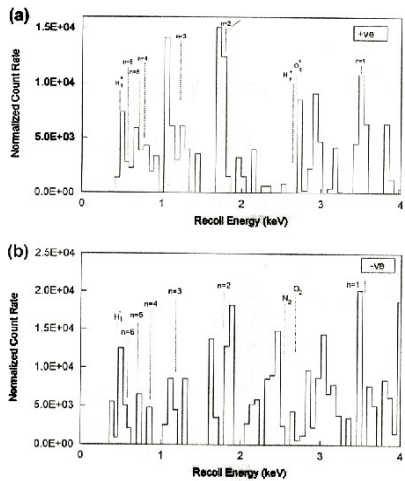

Fig. 5. The carbon DR spectra for $C^{n+}$ and $C^{n-}$ for $Ar^+ \rightarrow C$ at 150 keV.

## 4. Discussion

The data emerging out of the experimental results presented in Figs. 3 to 5 produces strong evidence in favor of the production of multiply charged carbon recoils. Both species i.e. positive as well as negative have been observed under energetic heavy ion bombardments. The energies obtained by monatomic recoils range from 1.45 keV for 100 keV $Kr^+$-C to 3.54 keV for 150 keV $Ar^+ \rightarrow C$. Our results clearly indicate that the probability of producing multiply charged recoil is high enough to produce $C^{n+}$ or $C^{n-}$ with $n \leq 6$. We have summarized some comparative features of recoils with charge number $n = 1$ and 2 of +ve and -ve charges in Table 1.

Table 1
Summary of some comparative features of recoils with charge number $n = 1$ and 2 of + ve and − ve charges.

|  | $E_0 = 100$ keV | | | 150 keV | | |
|---|---|---|---|---|---|---|
|  | $C^+:C^-$ | $C^+:C^{2+}$ | $C^-:C^{2-}$ | $C^+:C^-$ | $C^+:C^{2+}$ | $C^-:C^{2-}$ |
| $Ar^+$-C | 0.79 | 2.86 | 3.85 | 0.95 | 0.81 | 1.67 |
| $Kr^+$-C | 0.59 | 2.78 | 1.06 | 0.72 | 5.6 | 1.82 |

It has data for $Ar^+$ and $Kr^+$ at 100 and 150 keV. These values are taken from Figs. 2 to 5 for comparisons of respective peaks' counts/$\delta E$. The following points may be noted:

(I) The negatively charged species are produced in abundance compared with their positive counterparts except for 150 keV $Ar^+$;

(2) the trend of production of positive to negative recoils is increased for heavier projectiles i.e. $Ar^+ \rightarrow Kr^+$;

(3) the average probability of production of singly to doubly charged positive recoils increases for lighter projectiles;

(4) The $C^-:C^{2-}$ ratio shows the opposite trend and the average probability reduces with projectile's mass increase;

(5) The negatively charged spectra indicate that the peaks appear at positions slightly less than those for the positive recoils. The point will be further considered later while discussing the mechanisms of production of charged recoils.

In Table 2 the complete set of data for 100 and 150 keV $Ar^+ \rightarrow C$ and $Kr^+ \rightarrow C$ is presented. The data here is presented as ratios of $C^{1+}:C^{2+}:C^{3+}:C^{4+}:C^{5+}:C^{6+}$, : and similarly for the -ve species. In interpreting data in Table 2 we need to remember that $E_r$ has value of 1.45 keV for 100 keV $Kr^+ \rightarrow C$, 2.18 keV for 150 keV $Kr^+ \rightarrow C$, 2.36 for 100 keV $Ar^+ \rightarrow C$ and 3.54 keV for 150 keV $Kr^+ \rightarrow C$.

Table 2
The complete set of data for 100 and 150 keV Ar$^+$–C and Kr$^+$–C.

(a)

| $E_{DR}$ | C$^+$ | C$^{2+}$ | C$^{3+}$ | C$^{4+}$ | C$^{5+}$ | C$^{6+}$ |
|---|---|---|---|---|---|---|
| 1.45 | 1 | 0.36 | 0.23 | 0.51 | 0.07 | 0.06 |
| 2.18 | 1 | 0.18 | 0.68 | 0.06 | – | – |
| 2.36 | 1 | 0.35 | 0.4 | 0.16 | 0.14 | 0.20 |
| 3.54 | 1 | 1.24 | 0.93 | 0.43 | 0.52 | 0.62 |

(b)

| $E_{DR}$ | C$^-$ | C$^{2-}$ | C$^{3-}$ | C$^{4-}$ | C$^{5-}$ | C$^{6-}$ |
|---|---|---|---|---|---|---|
| 1.45 | 1 | 0.6 | 0.14 | 0.37 | – | 0.06 |
| 2.18 | 1 | 0.55 | 0.27 | 0.42 | – | – |
| 2.36 | 1 | 0.26 | 0.22 | 0.76 | 0 | 0.14 |
| 3.54 | 1 | 0.95 | 0.5 | 0.16 | 0.2 | 0.56 (?) |

$C^+ : C^-$ ratio has an average value of 0.76 ± 0.17 in this recoil energy range. The +ve to –ve ratio gradually rises with recoil energy Following the same pattern of comparison the ratio $C^{1+} : C^{2+}$ has an average of 0.3 ±0.12 at lower energies and increases for each higher recoil energy. The $n = +3$ charged fraction has the ratio 0.56 ±0.35 between recoil energies 1.45 to 2.36 with a two times increase at 3.54 keV. The $C^{4+}, C^{5+}, C^{6+}$, however, have non-consistent ratios mostly due to the difficulty in clearly distinguishing the overlapping peaks. The consistent behavior of $n \geq 3$ is that at $E_r \geq 2.36$, the production of higher charges is increased at least by a factor of 2.

We have observed that for 100 keV $Kr^+ \rightarrow C$ the $C^{n-}$ peaks appear at about 43 ± 2 eV less than the $C^{n+}$ peaks and similarly, the difference is 65 ± 5 eV for 150 keV case. This may indicate that perhaps the –ve recoils originate from collisions with surface carbon atoms and the positive ones are mostly due to subsurface scattered projectiles. One must remember that inward scattered projectiles produce recoils with $\theta_r > \theta_{DR}$ and hence with $E_r < E_{DR}$. In contrast, the positive recoils tend to be generated on and below the surface, because we have seen that $en+$ peaks have a high energy tail in most of the spectra. From this we expect $E_r > E_{DR}$ as $\theta_r < \theta_{DR}$ for majority of positive species.

In Figs. 3 and 4 there is evidence of $C_2$ ions being present with single and double charges. More work is needed to specifically look for this species.

## Conclusions

Spectra of multiply charged carbon recoils, positive as well as negative have been measured under energetic heavy ion bombardment of graphite. It has been observed that recoiling carbon atom seems to be easily willing to loose or accept electrons to form $C^{n\pm}$ ions where $n \geq 1$ fractions are not insignificant. The mechanisms of production of positive and negative species may, however, be different. Monatomic as well as dimers have been seen in these recoil energy spectra.